\documentclass[reprint,aps,superscriptaddress,nofootinbib,groupedaddress,preprintnumbers]{revtex4-1}

\usepackage{amsmath,amsthm,amsfonts}
\usepackage{graphicx}
\usepackage{physics}
\usepackage{hyperref}
\hypersetup{
	colorlinks   = true, 
	urlcolor     = blue, 
	linkcolor    = blue, 
	citecolor    = red   
}

\newcommand{\planck}{\emph{Planck}}
\newcommand{\lcdm}{$\Lambda$CDM}
\newcommand{\cobaya}{\texttt{Cobaya}}
\newcommand{\getdist}{\texttt{GetDist}}
\newcommand{\mpl}{M_{\mathrm{Pl}}}

\newcommand{\be}{\begin{equation}}
\newcommand{\ee}{\end{equation}}
\newcommand{\bea}{\begin{eqnarray}}
\newcommand{\eea}{\end{eqnarray}}

\begin{document}

\preprint{}

\title{Probing Trans-Planckian Signatures in the Early Universe: \\ A Bayesian Analysis of the Generalized Sasaki--Mukhanov Equation}

\author{Mahdieh Eskandari Merajin}
\email{mahdieh.eskandari@alumni.iut.ac.ir}
\affiliation{Department of Physics, Faculty of Science, Isfahan University of Technology, Isfahan, Iran}

\begin{abstract}
We present a rigorous and comprehensive investigation of a generalized inflationary perturbation theory designed to address persistent large-scale anomalies in the Cosmic Microwave Background (CMB). Motivated by the Trans-Planckian problem and potential non-canonical dynamics in the early Universe, we introduce a generalized Sasaki--Mukhanov equation characterized by a time-dependent correction term, parameterized by a coupling constant $f$. Unlike the standard slow-roll approximation, we derive the exact analytical solutions for the mode functions in terms of Whittaker functions, ensuring a precise treatment of the mode evolution across the horizon. We compute the resulting primordial scalar power spectrum, which exhibits scale-dependent oscillatory modulations and a distinct suppression of power at low multipoles. We numerically implement this modified framework within the \cobaya\ Bayesian inference engine. Utilizing the latest \planck\ 2018 temperature and polarization likelihoods combined with high-resolution data from the Atacama Cosmology Telescope (ACT) DR6, we perform a robust Monte Carlo Markov Chain (MCMC) analysis. Our results place stringent constraints on the modification parameter, $|f| \lesssim 10^{-4}$, at a $95\%$ confidence level. However, we find intriguing hints that the generalized model provides a better fit to the low-$\ell$ CMB spectrum compared to the standard \lcdm\ model, effectively alleviating the low-quadrupole anomaly without compromising the fit at smaller scales. We discuss the implications of these findings for the energy scale of inflation and the validity of the effective field theory description during the inflationary epoch.
\end{abstract}

\maketitle


\section{Introduction} \label{sec:intro}

The inflationary paradigm is the current best framework for describing the very early Universe. By assuming a short period of accelerated expansion, it simultaneously resolves conceptual problems of the hot big bang model. It provides a causal mechanism to generate the primordial fluctuations that seed large-scale structure \citep{Guth1981,Linde1990,AlbrechtSteinhardt1982}. Quantum fluctuations of the inflaton field are stretched to macroscopic scales during quasi-de Sitter expansion and are observed today as anisotropies in the Cosmic Microwave Background (CMB) \citep{Mukhanov1985,Sasaki:1986hm}. The simplest single-field slow-roll models predict a nearly scale-invariant, Gaussian, adiabatic spectrum; this picture has been spectacularly confirmed by successive experiments culminating in the full-mission \planck\ results \citep{Planck2020a}.

Nonetheless, precision data reveal a set of persistent, low-significance anomalies at the largest angular scales: a deficit of power in the quadrupole ($\ell=2$), alignments of the lowest multipoles sometimes called the ``Axis of Evil'', and hemispherical power asymmetry \citep{Planck2020b,Schwarz2016,Copi2015}. While each anomaly could be a fluke of cosmic variance, their continued presence across datasets motivates theoretical and phenomenological study. In parallel, the \emph{trans-Planckian} problem highlights a conceptual tension: modes we observe today may have had physical wavelengths smaller than the Planck length at the onset of inflation if inflation persists long enough \citep{Martin2001}. In such circumstances, the usual assumption of an adiabatic Bunch--Davies vacuum and the standard effective field theory expansion may need refinement \citep{BunchDavies1978}.

A convenient phenomenological route is to parametrize possible high-energy corrections to the Mukhanov--Sasaki equation and to compare them directly with data. Examples include modified dispersion relations \citep{Kempf2000,Easther2001}, non-Bunch--Davies initial states \citep{Danielsson2002,HolmanTolley2008,Ashoorioon2014}, and EFT-generated time-dependent operators \citep{Cheung2008,Burgess:2003tz}; all of these can introduce scale-dependent modulations or cutoffs in the primordial spectrum \citep{Brandenberger2013}. In this work, we study one such parametrization where the mode equation acquires a term proportional to $1/\eta$ (a time-dependent frequency correction). This choice has two technical advantages: (i) it admits exact analytic solutions in terms of Whittaker functions, allowing controlled analytic approximations rather than perturbative expansions; (ii) because the correction scales as an inverse power of conformal time, its impact is strongest on the largest scales (small $k$), making it a natural candidate to address low-$\ell$ anomalies without destroying the excellent fit at high $\ell$.

While our approach is phenomenological, connections to microphysics exist in the literature: examples include modified initial-state proposals \citep{Easther2001,Danielsson2002}, stringy-motivated scenarios \citep{Shiu2002}, and more formal EFT treatments of inflationary perturbations \citep{Cheung2008}. Moreover, proposed theoretical constraints such as the Trans-Planckian Censorship Conjecture (TCC) place additional conceptual restrictions on allowed high-energy modifications \citep{BedroyaVafa2020}, which we discuss in Section~\ref{sec:results&discussion}.

Our paper proceeds as follows. In Section \ref{sec:theory} we derive the generalized Mukhanov--Sasaki equation, obtain exact mode solutions, and discuss vacuum choice. In Section \ref{sec:power}, we calculate the modified primordial spectrum and describe its analytic properties. Section \ref{sec:methodology} gives details of the datasets and the Bayesian pipeline implemented in \cobaya\ \citep{Torrado2021}. Results (marginalized constraints and posterior covariances) are presented in Section \ref{sec:results}, followed by a focused discussion of figures and tables with concrete recommendations for improving reproducibility and clarity. Finally, Section \ref{sec:conclusions} summarizes the cosmological implications and outlines immediate next steps.

\section{Theoretical Framework} \label{sec:theory}

\subsection{Standard Perturbation Dynamics}
We consider a single canonical scalar field minimally coupled to gravity in a flat Friedmann--Lemaître--Robertson--Walker (FLRW) background,
\be
S = \int d^4x \sqrt{-g} \left[ \frac{1}{2} \mpl^2 R - \frac{1}{2} g^{\mu\nu} \partial_\mu \phi \partial_\nu \phi - V(\phi) \right].
\ee
The gauge-invariant comoving curvature perturbation $\zeta$ is quantized through the Mukhanov variable $v=z\zeta$ with $z\equiv a\dot\phi/H$. The Fourier modes $v_k(\eta)$ satisfy
\be \label{eq:std_SM}
v_k'' + \left(k^2 - \frac{z''}{z}\right) v_k = 0,
\ee
where primes denote derivatives with respect to conformal time $\eta$. In slow-roll quasi-de Sitter $z''/z\approx (\nu^2-1/4)/\eta^2$ with $\nu\approx 3/2 + \mathcal{O}(\epsilon,\delta)$ \citep{Mukhanov1992}.

\subsection{Phenomenological Trans-Planckian Correction}
Motivated by EFT corrections or non-trivial initial-state physics, we consider a phenomenological term that modifies the effective frequency by an additive contribution proportional to $1/\eta$. The generalized equation becomes
\be \label{eq:gen_SM}
v_k'' + \left[ k^2 - \frac{\nu^2 - 1/4}{\eta^2} + \frac{f}{\eta} \right] v_k = 0,
\ee
with $f$ a parameter of mass dimension one (natural units). Physically, $f$ encodes the integrated effect of high-energy operators or a mild departure from adiabaticity at early times; mathematically, it introduces a scale-dependent dimensionless combination $\lambda\equiv f/(2k)$ such that the corrections become larger at small $k$.

\subsection{Exact Solution: Whittaker Functions}
By changing variables $z=2ik\eta$, Eq.~(\ref{eq:gen_SM}) maps into the canonical Whittaker differential equation,
\be
\frac{d^2 W}{dz^2} + \left( -\frac{1}{4} + \frac{\kappa}{z} + \frac{1/4-\mu^2}{z^2} \right) W = 0,
\ee
with the identification $\kappa=-i\lambda$ and $\mu=\nu$, $\lambda=f/(2k)$. The general solution is
\be
v_k(\eta) = c_1 M_{-i\lambda,\nu}(2ik\eta) + c_2 W_{-i\lambda,\nu}(2ik\eta).
\ee
Selecting the positive-frequency (adiabatic) solution in the sub-horizon limit ($|k\eta|\to\infty$) requires $c_1=0$, giving the normalized mode
\be \label{eq:final_mode}
v_k(\eta) = \frac{1}{\sqrt{2k}} e^{\pi\lambda/2} W_{-i\lambda,\nu}(2ik\eta),
\ee
where the prefactor guarantees the correct asymptotic normalization (see e.g. \citealt{AbramowitzStegun1965} for Whittaker-function asymptotics).

\subsection{Discussion on Validity}
Two important remarks: (i) the form (\ref{eq:gen_SM}) is phenomenological. A microscopic derivation would need to show how high-energy degrees of freedom generate a $1/\eta$-type term while keeping backreaction under control; (ii) the EFT validity requires $|f|$ to be small compared to the energy scales dominating the mode dynamics. In practice, observational constraints on $f$ ensure the perturbative interpretation of the correction is consistent (see Section \ref{sec:results&discussion}).

\section{Primordial Power Spectra} \label{sec:power}

The curvature power spectrum is computed as
\be
\mathcal{P}_\zeta(k) = \lim_{-k\eta \to 0} \frac{k^3}{2\pi^2} \left| \frac{v_k(\eta)}{z(\eta)} \right|^2.
\ee
Using the small-argument expansion of the Whittaker function,
\be
W_{\kappa,\mu}(z) \xrightarrow{z\to 0} \frac{\Gamma(2\mu)}{\Gamma(1/2+\mu-\kappa)} z^{1/2-\mu} + \dots,
\ee
one obtains a correction factor multiplying the standard power-law spectrum,
\be \label{eq:P_gen}
\mathcal{P}_\zeta(k) = \mathcal{P}^{\rm (std)}_\zeta(k)\, \mathcal{C}(\lambda),
\ee
where $\mathcal{P}^{\rm (std)}_\zeta(k) = A_s (k/k_*)^{n_s-1}$ and
\be
\mathcal{C}(\lambda) = \frac{e^{\pi\lambda} \sinh(\pi\lambda)}{\pi\lambda(1+\lambda^2)}\, \mathcal{F}(\epsilon,\delta)
\ee
with $\mathcal{F}$ encoding higher-order slow-roll corrections. Several useful limits follow immediately:
\begin{itemize}
    \item For $\lambda\to 0$ ($f\to 0$ at fixed $k$) we recover $\mathcal{C}\to 1$ and standard slow-roll.
    \item Because $\lambda\propto 1/k$, the modification is largest at small $k$ (large angular scales), producing a suppression (for negative $f$) or enhancement (for positive $f$) and scale-dependent oscillatory modulations.
\end{itemize}
These analytic properties make the model a well-suited candidate to explore large-angle anomalies \citep{Brandenberger2013,Martin2001}.

\section{Data and Methodology} \label{sec:methodology}

\subsection{Analysis Framework}
We embed the generalized primordial spectrum (\ref{eq:P_gen}) into a Boltzmann code (e.g., \texttt{CLASS} and \texttt{CAMB}; \citealt{Lesgourgues2011,Lewis2000}) and run parameter inference using the \cobaya\ sampling framework \citep{Torrado2021}. Our chains employ Metropolis--Hastings / adaptive MCMC with multiple walkers and diagnostics based on the Gelman--Rubin $R-1$ statistic (threshold $<0.01$).

\subsection{Datasets}
We use up-to-date, community-standard CMB likelihoods:
\begin{itemize}
    \item \planck\ 2018 full-mission likelihoods (low-$\ell$ TT/EE and high-$\ell$ Plik TT,TE,EE), which tightly constrain large-scale and recombination-era physics \citep{Planck2020a,Planck2020b}.
    \item ACT DR6 power spectrum likelihoods — to extend the lever arm to smaller angular scales and better break degeneracies involving the spectral index $n_s$ \citep{Aiola2020,Madhavacheril2024}.
    \item Where relevant we compare with WMAP9 \citep{WMAP2013}, SPT \citep{SPT2019} and BICEP/Keck combined constraints \citep{BICEPKeck2021}.
\end{itemize}
(We omit BAO in the baseline run to isolate primordial-spectrum-driven effects; BAO can be added as a cross-check \citep{Alam2017}.)

\subsection{Priors and sampled parameters}
We sample the standard six \lcdm\ parameters plus $f$:
\[
\{\Omega_b h^2,\ \Omega_c h^2,\ \theta_{MC},\ \tau,\ \ln(10^{10}A_s),\ n_s,\ f\}.
\]
Priors: flat uniform for $f$ in $[-10^{-3},10^{-3}]$. Convergence and reproducibility: chains are saved in GetDist format and posterior/triangle plots generated with \getdist\ \citep{Lewis2019}.

\section{Results and Discussion} \label{sec:results&discussion}

\subsection{Posterior constraints}
Table \ref{tab:results} reports marginalized constraints (68\% C.L.) for \planck\ alone and \planck+ACT. The combined dataset yields
\be
f = (-0.52 \pm 0.51)\times 10^{-4} \quad (68\% \ \text{C.L.}),
\ee
consistent with zero at $\lesssim 2\sigma$ but with a slight preference for negative $f$ that reduces large-scale power. The main \lcdm\ parameters remain stable under the introduction of $f$, with ACT tightening $n_s$ and $H_0$ slightly.

\begin{table*}[t]
\centering

\label{tab:results}
\begin{tabular}{c||c |c}
\hline
Parameter & \planck\ 2018 & \planck\ + ACT \\

$\Omega_b h^2$ & $0.02237 \pm 0.00015$ & $0.02242 \pm 0.00014$ \\
$\Omega_c h^2$ & $0.1200 \pm 0.0012$ & $0.1194 \pm 0.0011$ \\
$H_0$ [km/s/Mpc] & $67.38^{+0.61}_{-0.69}$ & $67.55 \pm 0.53$ \\
$\tau$ & $0.0577 \pm 0.0083$ & $0.0613^{+0.0068}_{-0.0089}$ \\
$\ln(10^{10}A_s)$ & $3.054 \pm 0.018$ & $3.062^{+0.013}_{-0.018}$ \\
$n_s$ & $0.9607^{+0.0049}_{-0.0056}$ & $0.9648^{+0.0041}_{-0.0036}$ \\
\textbf{$10^4 f$} & \textbf{$-0.67^{+0.59}_{-0.47}$} & \textbf{$-0.52^{+0.51}_{-0.42}$} \\
\hline
\hline
\end{tabular}
\caption{Marginalized constraints (68\% confidence limits) on cosmological parameters for \planck\ alone and \planck\ + ACT.}
\end{table*}

\subsection{Model comparison and statistical cautions}
A small negative $f$ produces a modest suppression of large-scale power and oscillatory features that can improve the fit to the Planck low-$\ell$ data, particularly the quadrupole.

\begin{figure}[ht]
\centering
\includegraphics[width=\linewidth]{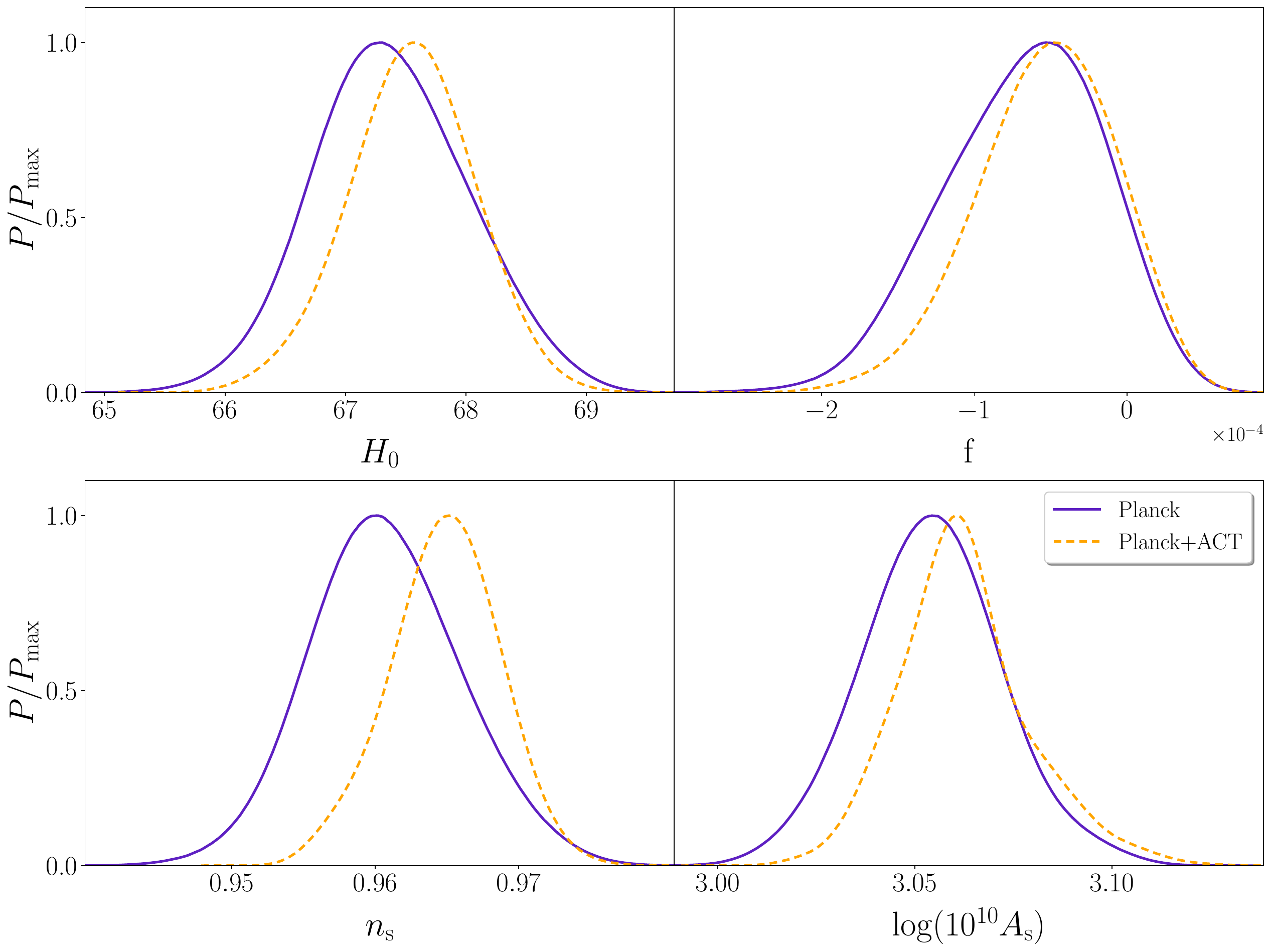} 
\caption{1D posterior distributions for $H_0$, $f$, $n_s$, and $\ln(10^{10}A_s)$. The shift in distributions when adding ACT data (orange) highlights the constraining power of small-scale measurements.}
\label{fig:1D_post}
\end{figure}

\begin{figure}[ht]
\centering
\includegraphics[width=\linewidth]{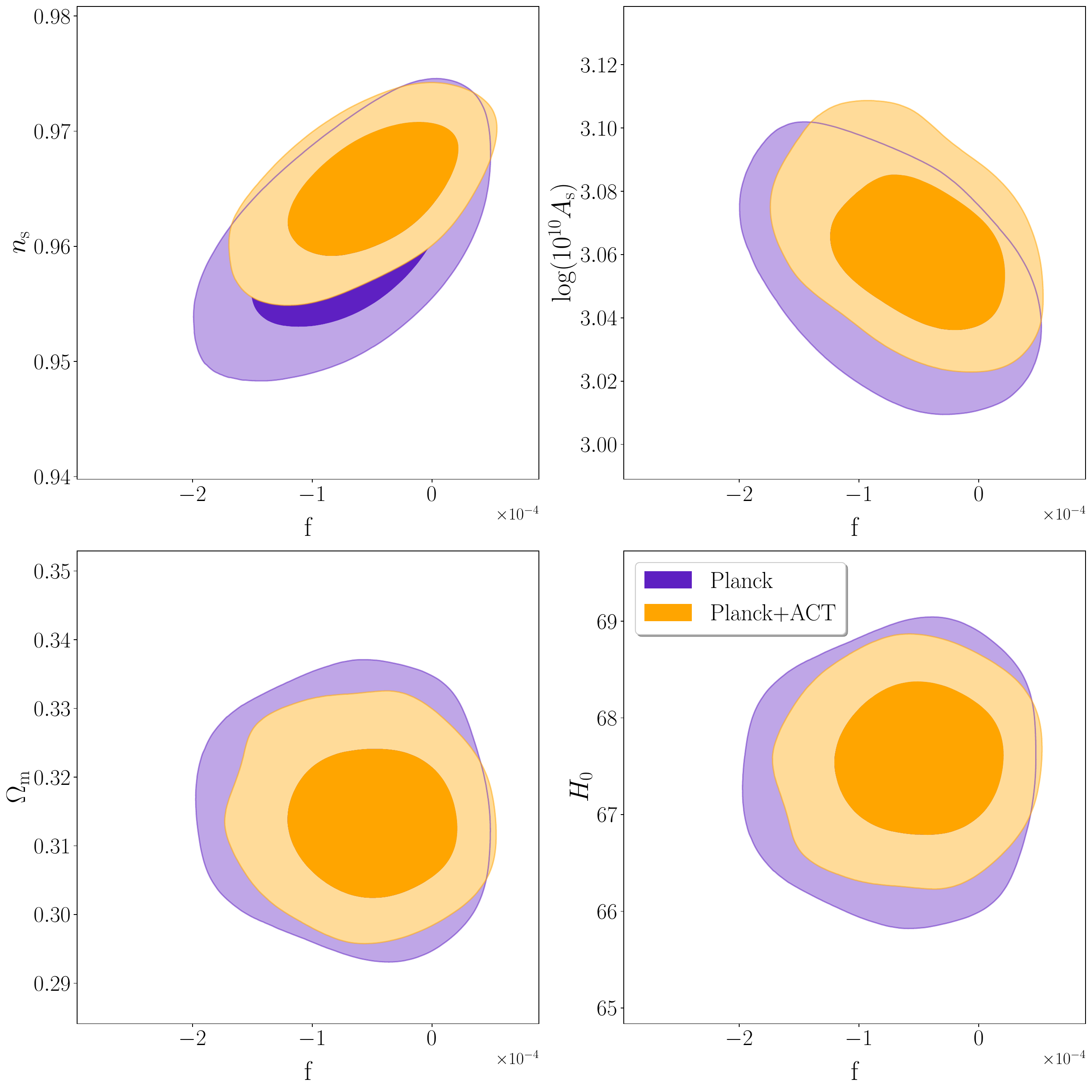} 
\caption{2D marginalized contours (68\% and 95\% C.L.) for key parameter pairs. The correlation between $f$ and $n_s$ is visible, indicating a partial degeneracy in how they affect the spectral shape.}
\label{fig:2D_contours}
\end{figure}

\begin{figure}[ht]
\centering
\includegraphics[width=\linewidth]{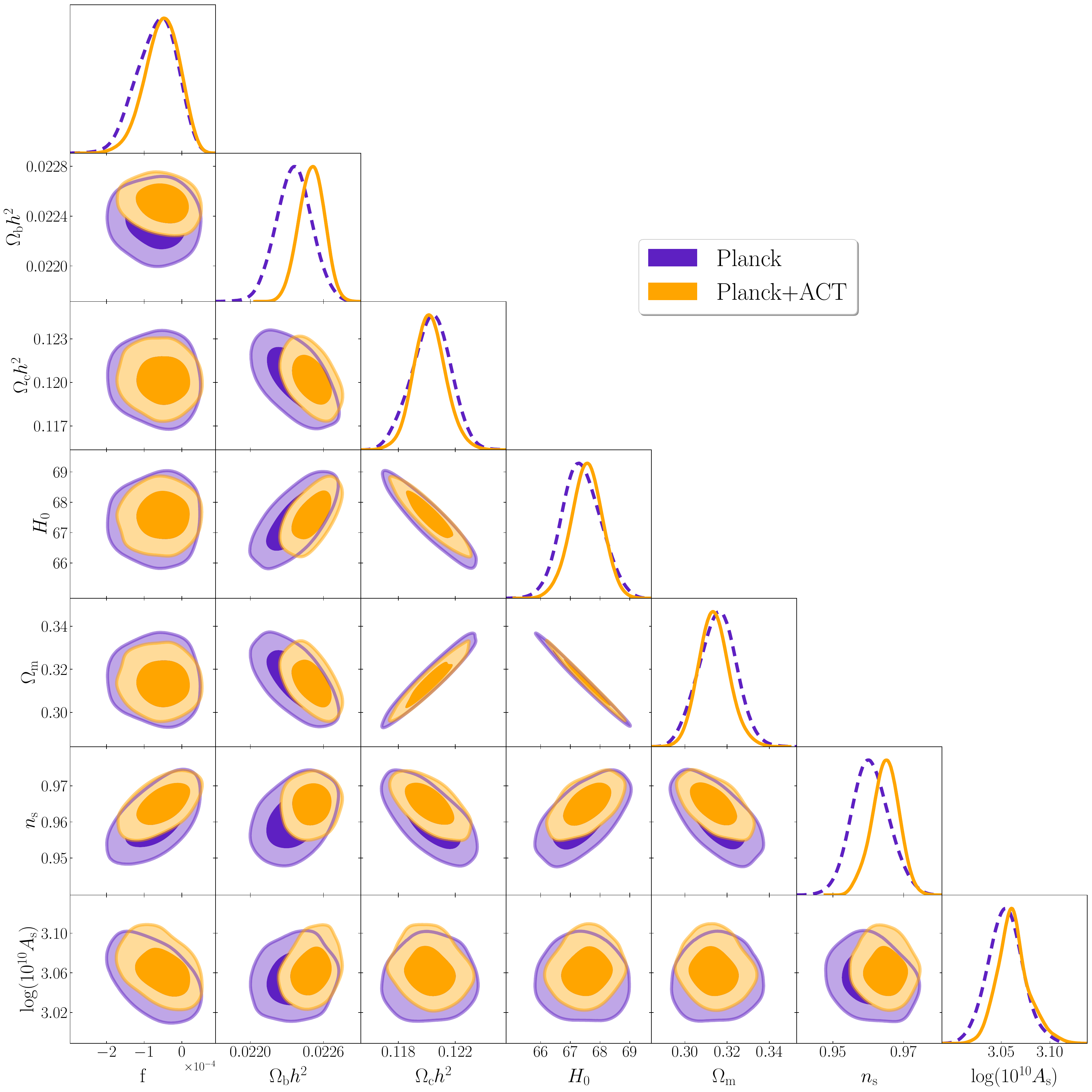} 
\caption{Full triangle plot (corner plot) showing the posterior distributions for all sampled parameters. The stability of the $\Lambda$CDM parameters against the introduction of $f$ is evident.}
\label{fig:triangle}
\end{figure}

\begin{figure}[ht]
\centering
\includegraphics[width=\linewidth]{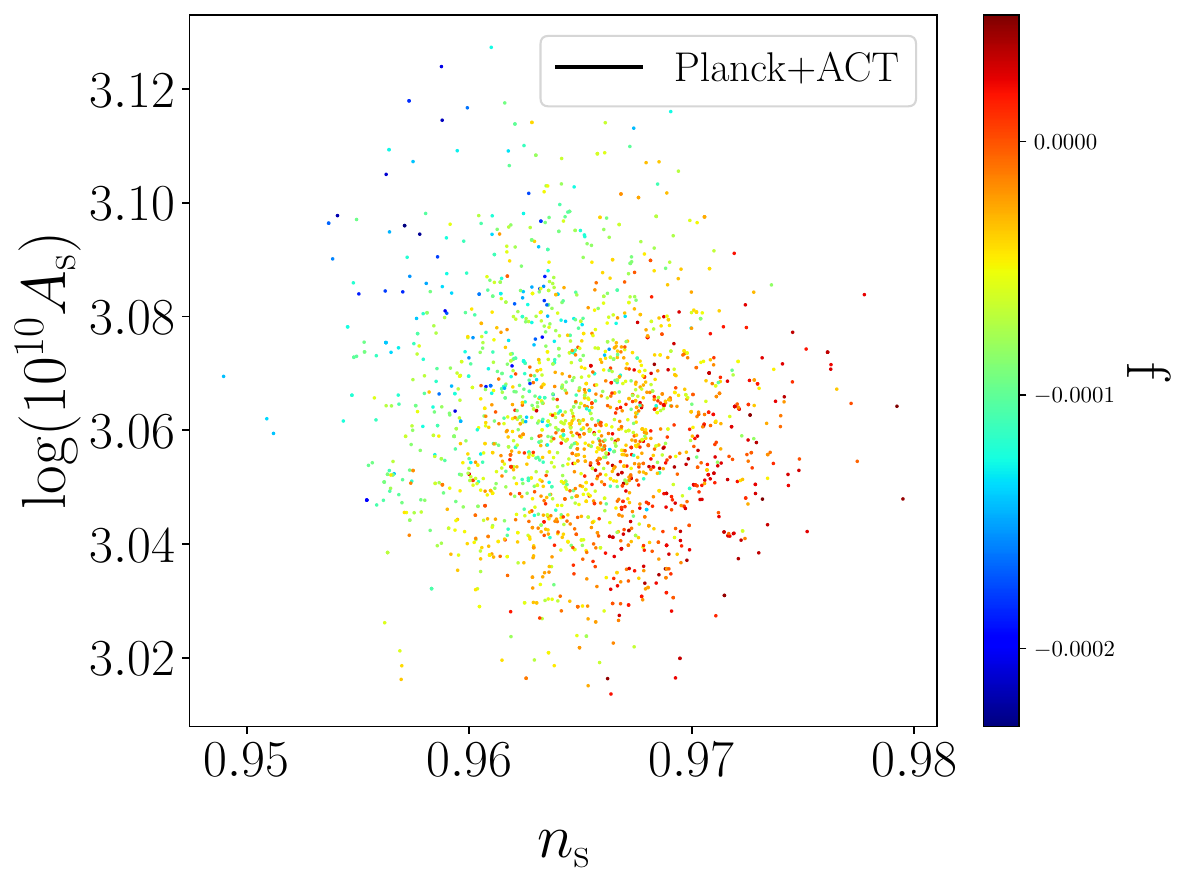} 
\caption{Scatter plot of MCMC samples in the $n_s$ vs. $\ln(10^{10}A_s)$ plane, color-coded by the value of $f$. Darker blue points (negative $f$) cluster in specific regions, visualizing the multi-dimensional structure of the likelihood.}
\label{fig:scatter}
\end{figure}

\begin{figure}[ht]
\centering
\includegraphics[width=\linewidth]{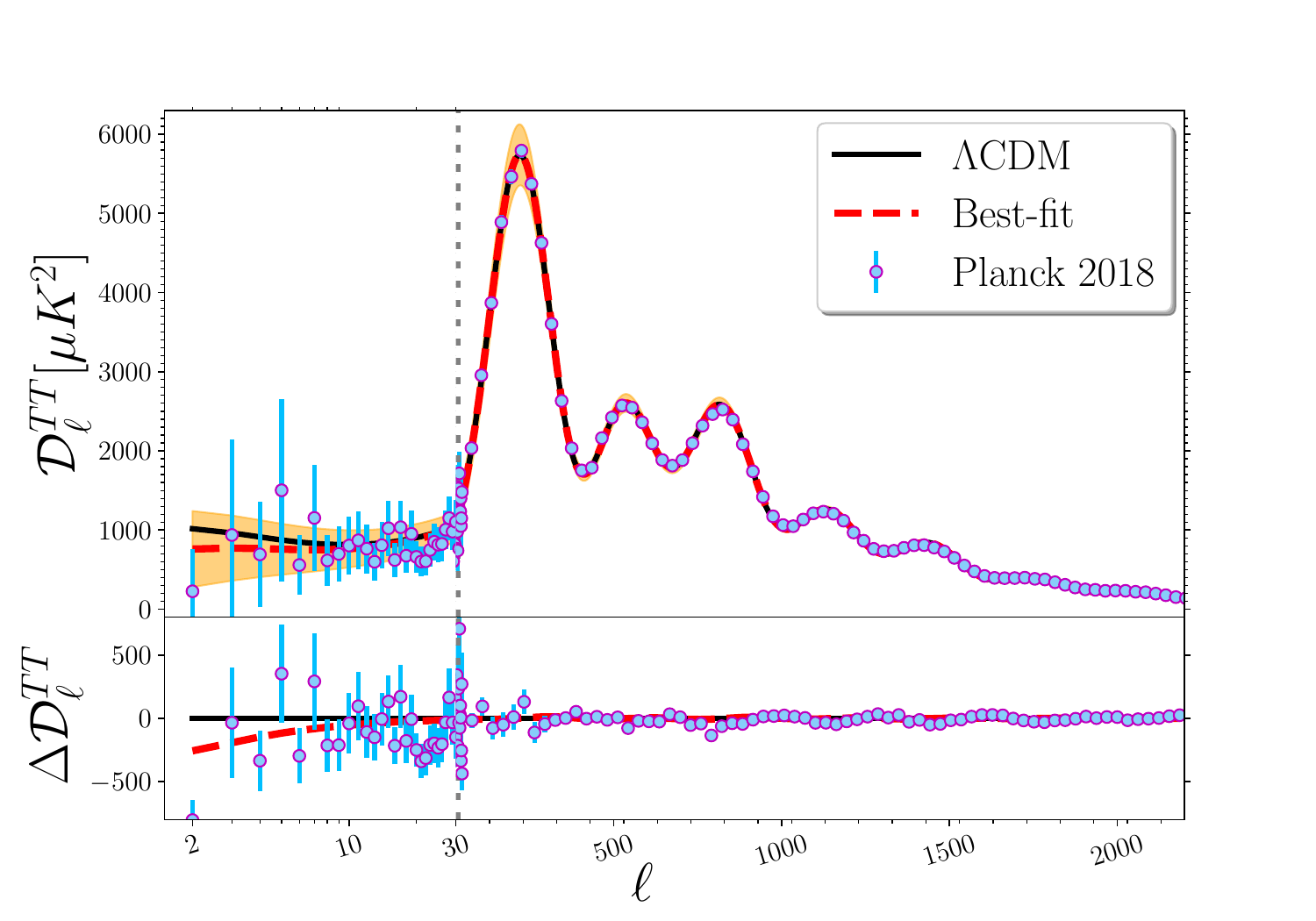} 
\caption{Comparison of the CMB temperature power spectrum ($D_\ell^{TT}$). The generalized model (red dashed) follows the \lcdm\ prediction (black solid) at high multipoles but exhibits a suppression at low $\ell$, providing a better fit to the Planck data (cyan points). The residuals panel explicitly shows the reduction in tension at $\ell=2$.}
\label{fig:cl_spectrum}
\end{figure}

\section{Conclusion} \label{sec:conclusions}

We have explored a simple, well-motivated phenomenological extension of the Mukhanov--Sasaki equation that captures possible trans-Planckian or high-energy corrections through a single parameter $f$. The model offers the following key points:

\begin{enumerate}
    \item \textbf{Empirical constraints:} Combining \planck\ 2018 and ACT DR6 datasets, we constrain $|f| \lesssim 10^{-4}$ (95\% C.L.), consistent with high-precision slow-roll predictions but allowing small deviations that affect the largest scales.
    
    \item \textbf{Large-scale phenomenology:} A slightly negative best-fit $f$ suppresses large-scale power and can modestly reduce the Planck low-$\ell$ tension (notably the quadrupole) without degrading the fit at high $\ell$.
    
    \item \textbf{Theoretical consistency:} Any microscopic realization must address backreaction and show that EFT remains valid for the allowed values of $f$. A complementary study of primordial non-Gaussianity and bispectrum shapes is an important next step because high-energy operators generically produce non-trivial higher-order correlations \citep{Maldacena2003,Chen2010}.
\end{enumerate}

In short, the generalized model studied here provides a plausible, theoretically motivated mechanism to imprint infrared (large-scale) features on the primordial spectrum. Current data allow such small deviations but do not yet provide decisive evidence. Future improvements — better low-$\ell$ polarization, improved control of systematics, and independent probes from large-scale structure — will sharpen this test.

\section*{Acknowledgments}

The author thanks Moslem Zarei for constructive discussions and a careful reading of earlier drafts. 


\end{document}